\documentclass[aps,prb,twocolumn]{revtex4}
\usepackage{graphicx,color}
\usepackage{amsmath}
\usepackage{amssymb,bm}

\newcommand{\f}[1]{Fig.~\ref{#1}}

\def\be{\begin{equation}}
\def\ee{\end{equation}}
\def\bea{\begin{eqnarray}}
\def\eea{\end{eqnarray}}
\def\l({\left(}
\def\r){\right)}

\begin{document}

\bibliographystyle{apsrev}
\title{Avalanche-driven fractal flux distributions in NbN superconducting films}

\author{I. A. Rudnev$^1$, D.~V. Shantsev$^{2,3}$,
T.~H.~Johansen$^{2,4,}$\cite{0}, A. E. Primenko$^5$}

\address{
$^1$ Moscow Engineering Physics Institute, 115409 Moscow, Russia\\
$^2$ Department of Physics, University of Oslo, P. O. Box 1048
Blindern, 0316 Oslo, Norway\\
$^3$ A. F. Ioffe Physico-Technical Institute, Polytekhnicheskaya
26, St.Petersburg 194021, Russia\\
$^4$ Texas Center for Superconductivity and Advanced Materials,
University of Houston, Houston, TX 77204 USA\\
$^5$ Department of Low Temperature Physics and Superconductivity,
Moscow State University, 117234 Russia}
\date{\today}

\begin{abstract}
Flux distributions in thin superconducting NbN films placed in a
perpendicular magnetic field have been studied using
magneto-optical imaging. Below 5.5~K the flux penetrates in the
form of abrupt avalanches resulting in dendritic structures.
Magnetization curves in this regime exhibit extremely noisy
behavior. Stability is restored both above a threshold temperature
$T^*$ and applied field $H^*$, where $H^*$ is smaller for
increasing field than during descent. The dendrite size and
morphology are strongly $T$ dependent, and fractal analysis of the
first dendrites entering into a virgin film shows that dendrites
formed at higher $T$ have larger fractal dimension.
\end{abstract}


\maketitle


Flux jumps are known to destroy the critical state of type-II
superconductors and suppress the apparent critical current
density.\cite{MR,GMR} In thin films the flux jumps manifest
themselves in a so-called dendritic instability, i.e.
avalanche-like penetration of magnetic flux along narrow branching
channels. Using magneto-optical (MO) imaging the dendritic
instability has been observed in superconducting films of
Nb,\cite{1967,duran,welling}
YBa$_2$Cu$_3$O$_{7-\delta}$\cite{leiderer,bolz03epl} (triggered by
a laser pulse), MgB$_2$\cite{epl,sust,apl,prb} and
YNi$_2$B$_2$C.\cite{jooss} Recently, flux dendrites were found
also in films of Nb$_3$Sn,\cite{rudnev} a superconductor with A15
structure widely used in applications.

In the present paper we report experiments made on niobium nitride
NbN films, another binary alloy shown here to have dendritic flux
penetration in the superconducting state. By combining MO imaging
and magnetometry we find threshold values for temperature and
applied field, and a striking critical behavior in the size and
morphology of the dendrites near the threshold.


Thin films of NbN were
fabricated by magnetron sputtering on sapphire substrates. Two
samples  with thickness 0.16 and 0.29~$\mu$m were selected for the
present studies.
 Table~1 shows their critical temperatures $T_c$,
the width of the superconducting transition $\Delta T_c$, the
critical current density $j_c$ at 4.2~K, and the resistivity
$\rho_N$ in the normal state. All these parameters were obtained
from transport measurements using a four-probe method.

\begin{table}
\begin{tabular}{|c|c|c|c|c|}
\hline
$d$ ($\mu$m) & $T_c$ (K) & $\Delta T_c$ (K) & $j_c$ (MA/cm$^2$) & $\rho_N$ ($\mu\Omega$m)\\
\hline
0.16 & 14.2 & 0.5 & 1.0 & 1.6 \\
\hline
0.29 & 15.0 & 0.5 & 1.4 & 1.1 \\
\hline
\end{tabular}
\caption{\label{tab1} Parameters of NbN films}
\end{table}


\begin{figure}[t]
\includegraphics [width=7.5cm]{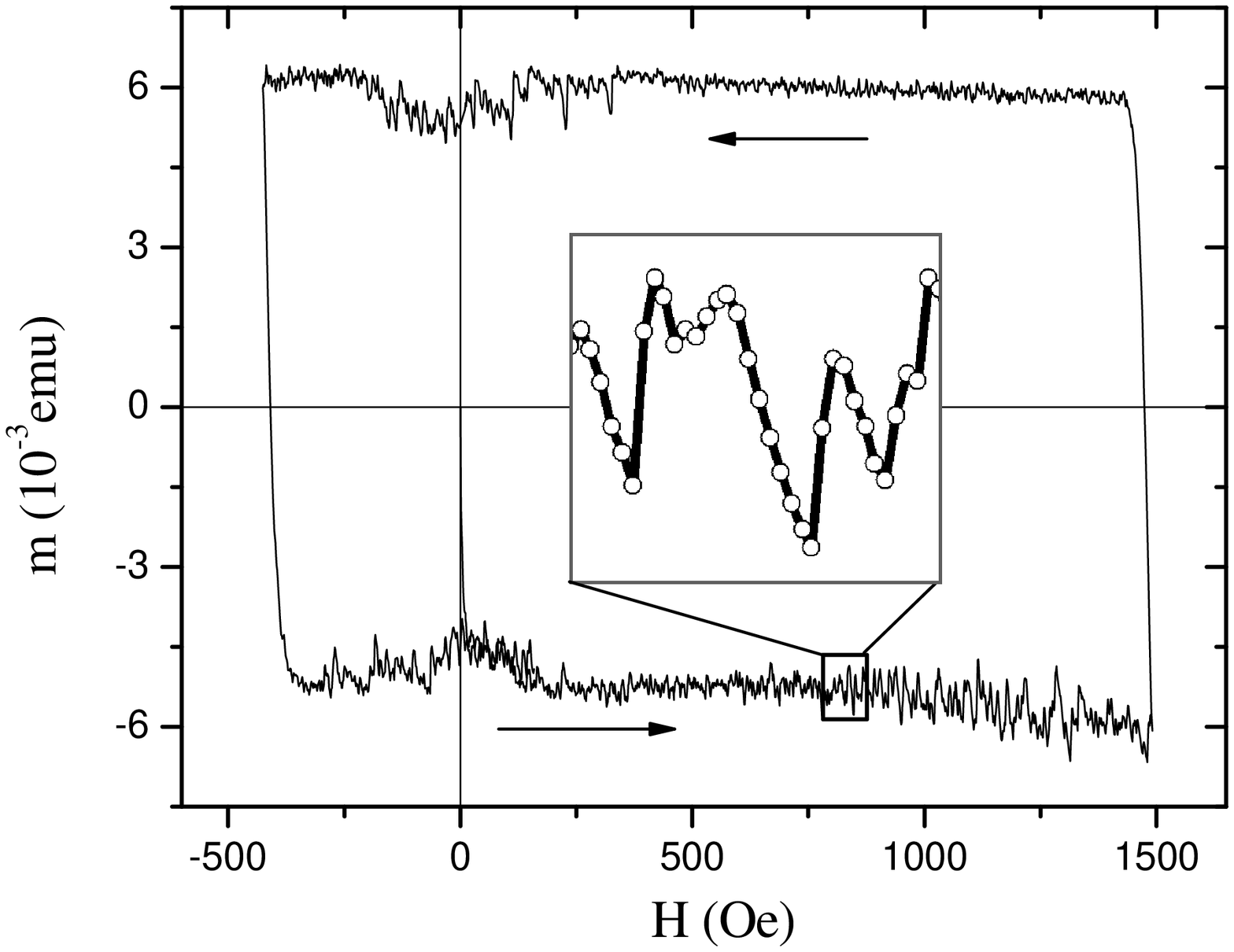}
\includegraphics [width=7.5cm]{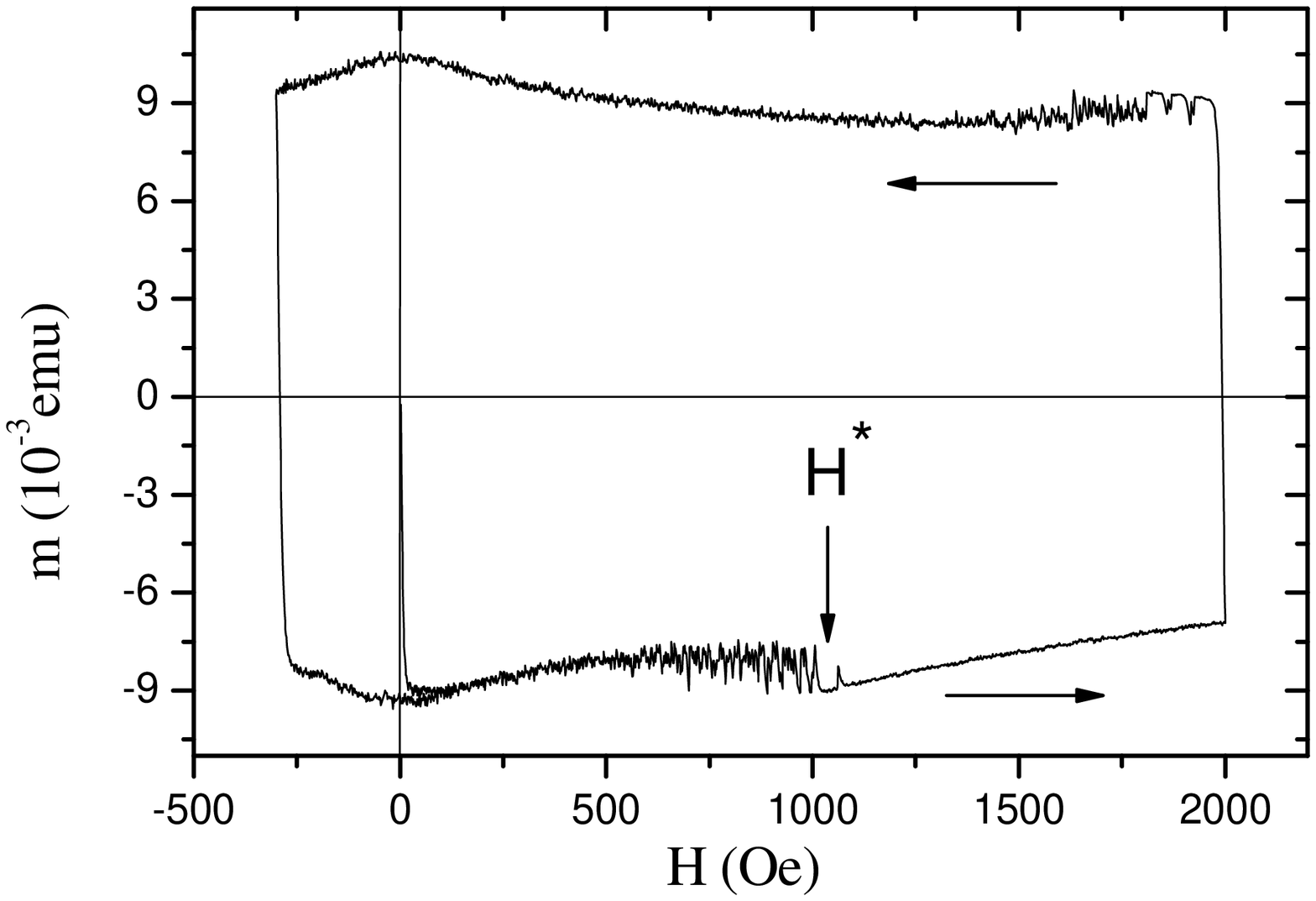}
\caption{\label{f_m} Magnetic moment as a function of applied
field at 4.2~K for two NbN films: 0.29~$\mu$m thick (upper) and
0.16~$\mu$m thick (lower).}
\end{figure}

The magnetic moment was measured using a vibrating sample
magnetometer PARC with a He flow cryostat. Shown in Fig.~1 are
magnetization curves $m(H)$ for the NbN films obtained at 4.2~K.
The instability manifests itself here in numerous and random jumps
of $m$. The typical jump amplitude is $\Delta m \sim 0.1m$, which
is much larger than the sensitivity of the magnetometer,
3$\cdot$10$^{-5}$~emu. Magnification of a part of the $m(H)$ curve
(see the inset) shows that the abrupt drops in magnetic moment are
followed by a much slower increase before the next drop occurs.

In the thinner film (lower panel) the jumps are
seen to disappear above a threshold field $H^* \approx 1$ kOe on
the increasing field branch. Moreover, just below $H^*$ the jumps
in $m$ have a larger amplitude and occur less frequently than at
low fields. On the descending field branch, the jumps reappear
when $m$ is fully reversed, and their amplitude decreases rapidly.

\begin{figure}[t]
\includegraphics [width=8.7cm]{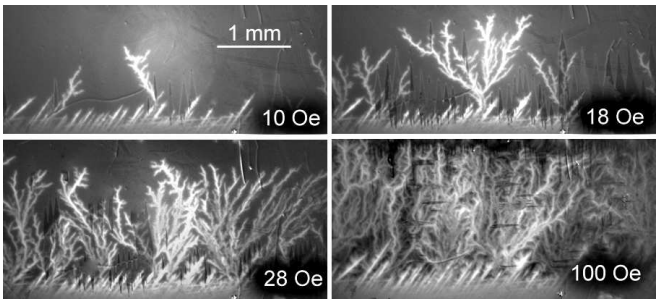}
\caption{\label{f_4im} Magneto-optical images of flux distribution
in a 0.29~$\mu$m thick NbN film at 3.5 K for increasing magnetic
field. The image brightness represents the magnitude of the local
flux density.}
\end{figure}

The nature
of these jumps in the magnetic moment was
clarified using MO imaging to visualize the dynamics of the full
flux distribution.\cite{MO} The sample, with a Faraday active
indicator film placed directly on top, was glued onto the cold
finger of an optical cryostat, where it was cooled to 3.5--8~K in
zero magnetic field (ZFC). Subsequently, a perpendicular field was
applied very slowly.

For low fields most part of the superconductor is in the Meissner
state and appear dark on the MO images. As the field increases,
the flux penetrates gradually, starting preferentially from the
weak places along the edges.
At some field $H_\text{fj}\approx 10$~Oe an abrupt invasion of a
relatively large flux structure occurred, see \f{f_4im}. Further
field increase resulted in formation of even larger and highly
dendritic structures entering one by one. Eventually, when
reaching $H \approx 28$~Oe, the flux dendrites filled most of the
film area. Upon further field increase, new dendrites continued to
form, but now entering on top of already existing ones, as seen in
the MO image taken at 100~Oe. Note that the tilted lines of flux
penetration near the edge is of a different origin, most likely
due to polishing streaks in the substrate.


Most features of the observed dendritic instability in the NbN
films resemble those found previously in other
materials.\cite{1967,duran,welling,leiderer,bolz03epl,epl,sust,apl,prb,jooss,rudnev,other}
The dendrites propagate into the film faster than 1~ms, which is
the time resolution of the CCD camera recording our MO images. In
fact, we expect that the propagation is even much faster, as was
found using ultrafast MO imaging of dendrites in
YBa$_2$Cu$_3$O$_{7-\delta}$ and MgB$_2$
films.\cite{bolz03epl,privcomm}
Another characteristic feature is that once a dendritic structure
is formed, it remains ``frozen'' and does not grow any further
during subsequent increase of the applied field. Moreover, when
the experiment is repeated under identical conditions, the exact
field when the dendrites form and their exact pattern are never
repeated. In the present films the dendritic instability was
observed only below $T^*=5.5$~K, whereas a similar threshold
temperature in MgB$_2$ is 10~K.\cite{epl} Above these threshold
temperatures the flux penetration is always spatially smooth and
gradual in time.

The instability disappears not only when $T>T^*$, but also when
the field becomes sufficiently high, $H>H^*$, see \f{f_m}(lower
panel).  Our results clearly show
that the threshold value depends on the field sweep direction. We
propose that this dependence originates from vortex annihilation,
which takes place only for the decreasing field case. Indeed, when
$H$ is increasing, the screening currents generate near the film
edge a strong demagnetization field of the same sign as $H$.
However, for decreasing $H$, the direction of screening currents
and demagnetization field changes to the opposite. As a result,
the film edge experiences a negative external field that also
partly penetrates inside. In this state, there is a line near the
edge where vortices and antivortices meet.\cite{BrIn,zeld} Their
annihilation releases additional energy that can facilitate the
triggering of the instability.\cite{beasley} Consequently, one may
expect that the dendritic instability occurs in a wider range of
applied fields $H$ along the descending field branch as compared
to the ascending branch.

\begin{figure}
\includegraphics [width=8.7cm]{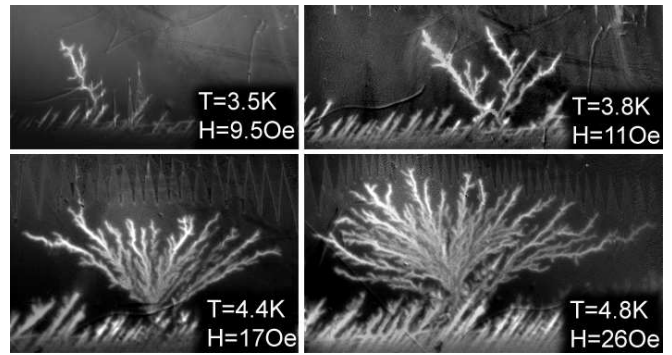}
\caption{\label{f_first} First dendrites formed in a zero-field-cooled NbN film
during increasing applied field $H$.
Dendrites formed at higher temperatures are characterized by a stronger branching.
}
\end{figure}

\begin{figure}
\includegraphics [width=8cm]{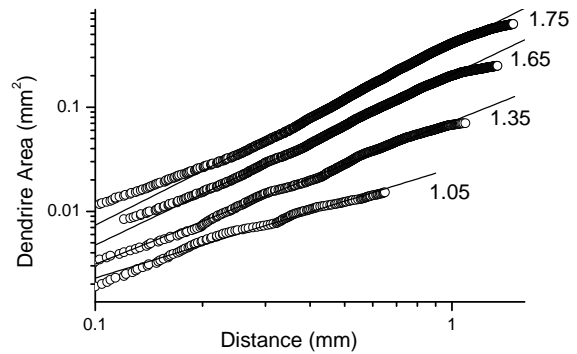}
\caption{\label{f_d} Determination of fractal dimension of the first flux dendrite
for different temperatures: 3.5, 3.8, 4.4 and 4.8~K from top to bottom. The curves are
shifted vertically to avoid overlapping.}
\end{figure}

\begin{figure}
\includegraphics [width=8cm]{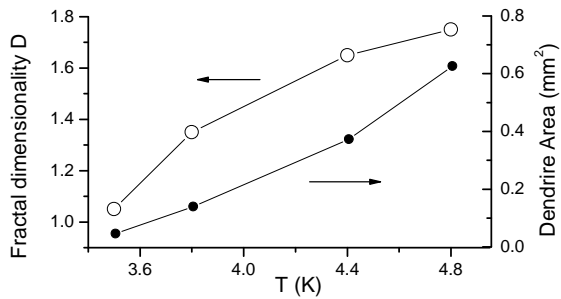}
\caption{\label{f_da} Fractal dimension and total area of the first flux dendrite
as functions of temperature.}
\end{figure}

The existence of a threshold field $H^*$ was reported earlier in
the magnetization studies of MgB$_2$ films.\cite{zhao} Our
observed asymmetry for the increasing and decreasing field sweeps
is also in agreement with results of Ref.~\onlinecite{jooss},
where dendritic jumps were found only for decreasing $H$.
Interestingly, during the dendrite growth the annihilation zone
may propagate very deep into the film. This is confirmed by
observation of ``negative'' flux in the dendrite core that
propagated into a film containing positive flux for a decreasing
$H$.\cite{sust}

Figure~\ref{f_first} shows MO images of the very first dendrite
formed in the 0.29~$\mu$m thick ZFC film during 4 experiments at
slightly different temperatures. These dendrites were formed also
at different (first jump) fields $H_\text{fj}$, and there is a
clear tendency that $H_\text{fj}$ increases with temperature.
Dendrites formed at higher $T$ and $H_\text{fj}$ are also larger
in size and more branching, a tendency can be traced all the way
up to $T^*=5.5$~K.  Note also that the $m(H)$ curve in
\f{f_m}(lower panel) exhibits increasingly larger jumps as the
threshold field $H^*$ is approached. Therefore, it is a general
trend that the dendritic structures have maximal size when the
system is close to the stability limit, i.e. for $H\approx H^*$ or
$T\approx T^*$.

To quantify these changes in morphology
of the branching flux structures we made a fractal analysis of
their shape. The MO images were discretized to obtain a cluster of
pixels constituting a dendrite structure.
Then we calculated the number of pixels $N(R)$ that fall inside a
circle of radius $R$ with the center at the dendrite root. If the
structure gives a power-law, $N \propto R^D$, the exponent gives
the fractal dimension $D$ of the cluster.\cite{feder}

The results of this analysis is presented in \f{f_d}, where the
actual dendrite area represents $N$. We find a reasonably good
power-law behavior, and the fractal dimension increases with
temperature, see \f{f_da} for a summary. The dimension
changes from approximately unity at the lowest $T$ to $D =1.75$
for the most branching structure at 4.8~K.
Note that in general,
the dendritic fingers do not exceed the middle of the strip,
where the Meissner currents (and hence Lorentz force) change to
the opposite direction. This limits the size of all dendritic
structures. As a result, more branching dendrites with larger $D$
always have a larger area.

A similar temperature dependence of dendrite morphology has been
reported earlier for Nb\cite{duran,welling} and for
MgB$_2$\cite{epl} samples. Various degrees of branching have also
been obtained by simulations taking into account the heat produced
by flux motion,\cite{epl,aranson04} which suggests
a thermal origin of the instability. Our present results (i) give
a quantitative measure of the branching, $D$, and (ii) show that
simultaneous increase of $D$ and dendrite area is a more general
effect since it takes place when approaching the instability
threshold $H^*(T)$ either by changing $T$ or $H$.


This work is supported by FUNMAT\verb+@+UiO, and The Research
Council of Norway.


\end{document}